\documentclass[%
 reprint,
 superscriptaddress,
 showkeys,
nofootinbib,
 amsmath,amssymb,
 aps,
]{revtex4-2}

\usepackage{soul}
\usepackage{float}
\usepackage[caption=false]{subfig} 

\usepackage{graphicx}
\usepackage{dcolumn}
\usepackage{bm}
\usepackage[hypertexnames=true,colorlinks=true,urlcolor=blue,bookmarks=true,citecolor=blue,breaklinks=true,pdftex]{hyperref}

\usepackage{array}
\usepackage{multirow}
\newcolumntype{C}[1]{>{\centering\arraybackslash}p{#1}}

\begin{document}

\preprint{APS/123-QED}

\title{Evaluating Machine Learning Models for Supernova Gravitational Wave Signal Classification}

\author{Y. Sultan Abylkairov}
\email{sultan.abylkairov@nu.edu.kz}
\affiliation{Department of Mathematics, Nazarbayev University, 010000 Astana, Kazakhstan}
\affiliation{Energetic Cosmos Laboratory, Nazarbayev University, 010000 Astana, Kazakhstan}
\author{Matthew C. Edwards}
\affiliation{Department of Statistics, University of Auckland, Auckland 1010, New Zealand}
\author{Daniil Orel}
\affiliation{Department of Natural Language Processing, Mohamed bin Zayed University of Artificial Intelligence, Abu Dhabi 54115, United Arab Emirates}
\author{Ayan Mitra}
\affiliation{Center for Astrophysical Surveys, National Center for Supercomputing Applications, University of Illinois Urbana-Champaign, Urbana, IL, 61801, USA}
\affiliation{Department of Astronomy, University of Illinois at Urbana-Champaign, Urbana, IL, 61801, USA}
\author{Bekdaulet Shukirgaliyev}
\affiliation{Heriot-Watt University Aktobe Campus, 030000 Aktobe, Kazakhstan}
\affiliation{K. Zhubanov Aktobe Regional University, 030000 Aktobe, Kazakhstan}
\affiliation{Department of Physics, Nazarbayev University, 010000 Astana, Kazakhstan}
\author{Ernazar Abdikamalov}
\affiliation{Department of Physics, Nazarbayev University, 010000 Astana, Kazakhstan}
\affiliation{Energetic Cosmos Laboratory, Nazarbayev University, 010000 Astana, Kazakhstan}

\date{\today}

\begin{abstract}
We investigate the potential of using gravitational wave (GW) signals from rotating core-collapse supernovae to probe the equation of state (EOS) of nuclear matter. By generating GW signals from simulations with various EOSs, we train machine learning models to classify them and evaluate their performance. Our study builds on previous work by examining how different machine learning models, parameters, and data preprocessing techniques impact classification accuracy. We test convolutional and recurrent neural networks, as well as six classical algorithms: random forest, support vector machines, na\"{i}ve Bayes, logistic regression, $k$-nearest neighbors, and eXtreme gradient boosting. All models, except na\"{i}ve Bayes, achieve over 90 per cent accuracy on our dataset. Additionally, we assess the impact of approximating the GW signal using the general relativistic effective potential (GREP) on EOS classification. We find that models trained on GREP data exhibit low classification accuracy. However, normalizing time by the peak signal frequency, which partially compensates for the absence of the time dilation effect in GREP, leads to a notable improvement in accuracy. Despite this, the accuracy does not exceed 70 per cent, suggesting that GREP lacks the precision necessary for EOS classification. Finally, our study has several limitations, including the omission of detector noise and the focus on a single progenitor mass model, which will be addressed in future works. 
\end{abstract}

                               
\maketitle

\section{Introduction}
\label{sec:intro}

Gravitational waves (GWs) are ripples in the fabric of space-time that travel at the speed of light. GWs can traverse the cosmos without significant disturbance, reaching Earth mostly unchanged and providing a clear view of their sources. To date, GW detections have come from black hole mergers \cite{abbott2016gw150914}, neutron star mergers \cite{abbott2017gw170817}, and black hole-neutron star mergers \cite{abbott2021observation}, offering unprecedented insights into these phenomena. Another promising source of GWs is core-collapse supernovae (CCSNe), which could greatly enhance our understanding of these events \cite{kotake:13review, Hayama18Circular, schneider19equation, abbott2020optically, szczepanczyk2023optically, powell22inferring}. Current GW observatories can detect supernova signals from within our Galaxy, expected once or twice per century \cite{gossan16observing, adams13observing}. Future detectors will have higher sensitivity, enabling them to observe more distant events \cite{srivastava19detection}.

CCSNe are powerful explosions that mark the death of massive stars. As these stars age, they undergo successive stages of nuclear fusion, ultimately forming a dense core of iron-group nuclei. Once the core reaches large enough mass, the electron degeneracy pressure can no longer support it against gravity, causing it to collapse. The collapse is abruptly halted upon reaching nuclear densities, triggering a rebound that generates a shock wave. For a supernova explosion to occur, the shock must expel the star's outer layers, leaving behind a stable neutron star. Otherwise, a black hole forms \cite{oconnor11, cerda:13, Burrows23Black}. The precise mechanisms behind this process is a focus of ongoing research \cite[][for recent reviews]{janka16physics, burrows13colloquium, muller20hydrodynamics, mezzacappa20physical}. 

Neutrinos play a key role in CCSNe. The protoneutron star (PNS) cools by emitting vast numbers of neutrinos \cite{oconnor13progenitor,mueller:14}. A small fraction of them are absorbed behind the shock, depositing their energy and heating the region \cite{Lentz12Interplay, mueller:12a}. This heating induces convection \cite{herant94inside, burrows95on, janka95first}, which, along with the potential development of standing accretion shock instability (SASI) \cite{blondin03stability, foglizzo06neutrino}, helps push the shock front outward, driving the explosion. This mechanism is known as the neutino mechanism of CCSNe.

In rare cases, stars are rapidly rotating and the PNSs are born with substantial rotational kinetic energy \cite{burrows:07b, winteler12, moesta:14b, obergaulinger:20, kuroda:20}. Magnetic fields can transfer this energy to the shock, potentially causing more powerful hypernova explosions \cite{woosley_bloom:06} and possibly triggering long gamma-ray bursts \cite{woosley:06, metzger11protomagnetar}. This is commonly referred to as the magnetorotational mechanism of CCSNe.

The supernova GWs are predominantly emitted by the PNS dynamics \cite[][for recent reviews]{abdikamalov22gravitational, mezzacappa24gravitational}. Neutrino-driven convection and SASI perturb the flow, driving PNS oscillations \cite{Murphy09Model, radice:19gw, Andresen17Gravitational, Mezzacappa23Core, Vartanyan23Gravitational}. In rapidly rotating stars, the centrifugal force causes the collapse to become deformed, resulting in a core bounce with quadrupolar deformation. This excites oscillations of the PNS that last for $\sim 10 $ ms after bounce \cite{ott12correlated}. This type of signal is usually called the rotating bounce GW signal. 

Gravitational waves originating from the PNS oscillations carry valuable information about its structure \cite{mueller:13, Morozova18GW, radice:19gw, pajkos19, pajkos21, Powell24Determining}. In particular, the signal depends on the poorly understood properties of high-density nuclear matter \cite{richers:17, schneider19equation, Wolfe23GW, Powell24GW}. This offers an opportunity to constrain the parameters of the nuclear equation of state (EOS) using GW data \cite{edwards2017, chao22determining}.

Recently, machine learning (ML) has shown significant promise in inferring source parameters from GW signals \cite{edwards2017, Astone18New, chan2020, Iess_2020, Lopez21, antelis22using,  saiz-perez22classification, mitra23, Eccleston24generative, morales2024residualneuralnetworksclassify, nunes2024deep}. It has proven particularly effective for rotating bounce signals, which are easier to model and cost-efficient to simulate, allowing the generation of large GW datasets necessary for ML training \cite{mitra24}. Supernova detection pipelines have been built using convolutional neural networks (CNNs) \cite{Astone18New, Iess_2020, faisal:2024}, the pre-trainied Mini-Inception ResNet \cite{Lopez21}, and long short-term memory (LSTM) networks \cite{iess:2023}.  Linear discriminant analysis (LDA) and support vector machines (SVMs) have been used to enhance gravitational-wave follow-up for core-collapse supernovae \cite{antelis22using}.  A residual network model has been used to classify high frequency emission in core-collapse supernovae \cite{morales2024residualneuralnetworksclassify}.  Classifying the supernova explosion mechanism has been performed using CNNs \cite{chan2020}, dictionary learning \cite{saiz-perez22classification}, and both CNNs and dictionary learning \cite{Powell24Determining}. Inference on continuous parameters has also been performed; using CNNs and transfer learning to infer rotation rate \cite{chao22determining}, and using residual CNNs and principal component analysis (PCA) to infer $D\cdot\Delta h$ (i.e., the distance multiplied by the difference between the maximum and minimum strain during bounce) as well as $f_{\text{peak}}$ (i.e., the peak frequency after bounce) \cite{nunes2024deep}.  A generative adversarial network (GAN) has been used to learn and rapidly emulate core-collapse signals, useful as a phenomenological model and for data augmentation \cite{Eccleston24generative}. In \cite{edwards2017, chao22determining}, CNNs were applied to EOS classification using the GW database from \cite{richers:17}. Mitra et al. \cite{mitra24} extended this work by incorporating uncertainties in electron capture rates during collapse, which influence the dynamics and, consequently, the GW signal. We show that the enhanced dataset improves classification accuracy by providing a larger volume of data, even though it includes different electron capture rates. Additionally, we demonstrate that the highest classification accuracy is achieved for models with moderate rotation rates: neither too rapid nor too slow. This suggests that moderate rotation offers the optimal balance: it flattens the core sufficiently to generate a significant quadrupole moment change (and thus the GW signal), while avoiding excessive centrifugal support that would hinder the core from reaching the high densities where differences between various EOSs become most pronounced.

In this work, we expand on previous studies in two key ways. First, we investigate how classification accuracy is affected by different ML model types, parameters, and data preprocessing techniques. Second, we examine the impact of a GW signal approximation based on the general relativistic effective potential (GREP) \cite{Rampp02Radiation, marek:06, bmueller:08}, which modifies the gravitational potential to approximate general relativistic (GR) effects within a Newtonian framework. The main advantage of this method, which is widely used for CCSN simulations \citep{oconnor18two, pajkos19, radice:19gw, Bruenn20CHIMERA, obergaulinger:20, Bugli21Three, Nakamura22Three, varma23}, is its simplicity and lower computational cost compared to full GR simulations. However, the main drawback is that while GREP can mimic GR gravity, it does not capture other relativistic effects, such as time dilation or length contraction. We assess how well ML models trained on the GW signals generated using the GREP approximation can classify the EOS from the rotational bounce GW signal.

This paper organized as following. In Section \ref{sec:method}, we describe our dataset, ML methods, and accuracy evaluation criteria. In Section \ref{sec:results}, we present our results. Finally, in Section \ref{sec:concl}, we summarize our results and provide conclusions.

\section{Methodology}
\label{sec:method}

In this section, we describe the dataset, the ML algorithms, hyperparameter selection, training process, and the evaluation method for EOS classification.

\subsection{Data}
\label{sec:data}

We obtain GWs from numerical simulations using the code {\tt CoCoNuT} \cite{Dimmelmeier02a, dimmelmeier:05MdM}. We perform two sets of simulations. In the first set, we use GR with the conformal-flatness condition (CFC) \cite{isenberg:08, wilson96, cordero-carrion:09}, which yields excellent accuracy in the context of CCSNe \cite[e.g.,][]{cerda05, shibata:04, ott:07cqg}. In the second set, we use Newtonian hydrodynamics with GR effective potential. We use the so-called ``case A" formulation, which was found to better reproduce GR results \citep{marek:06} (see Appendix~\ref{sec:grep} for details). 

\begin{figure*}[t]
    \centering
    \subfloat{{\includegraphics[width=.47\textwidth]{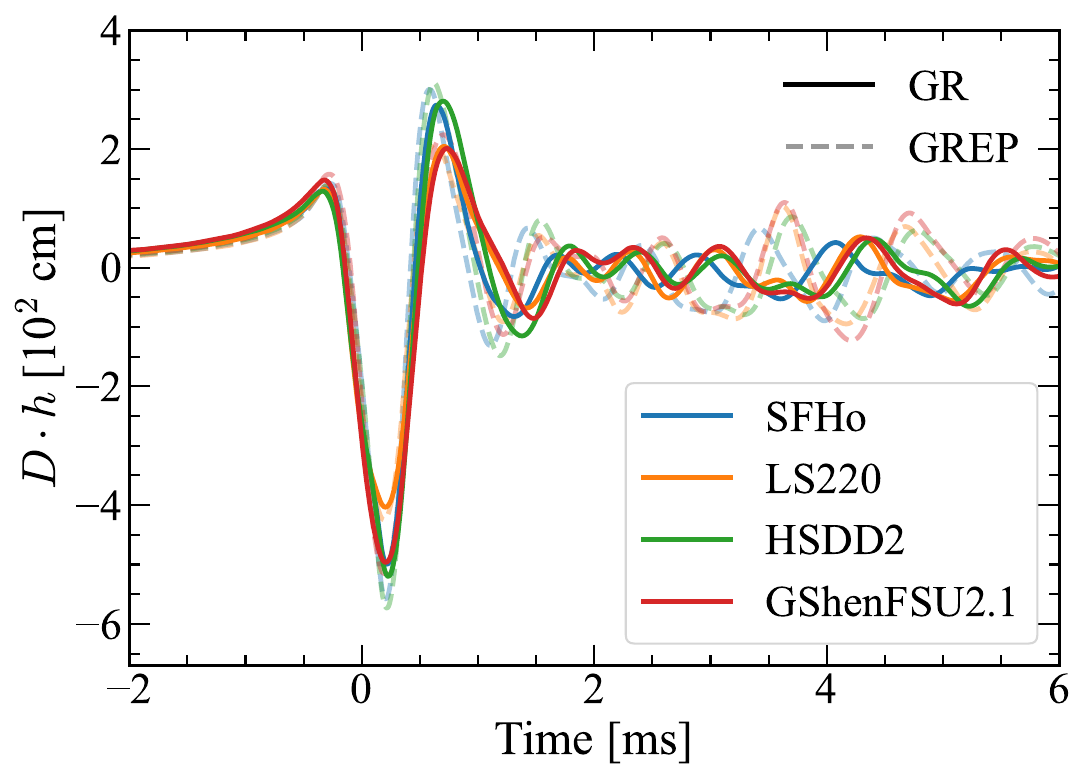} }}
    \hspace{0.03\textwidth}
    \subfloat{{\includegraphics[width=.47\textwidth]{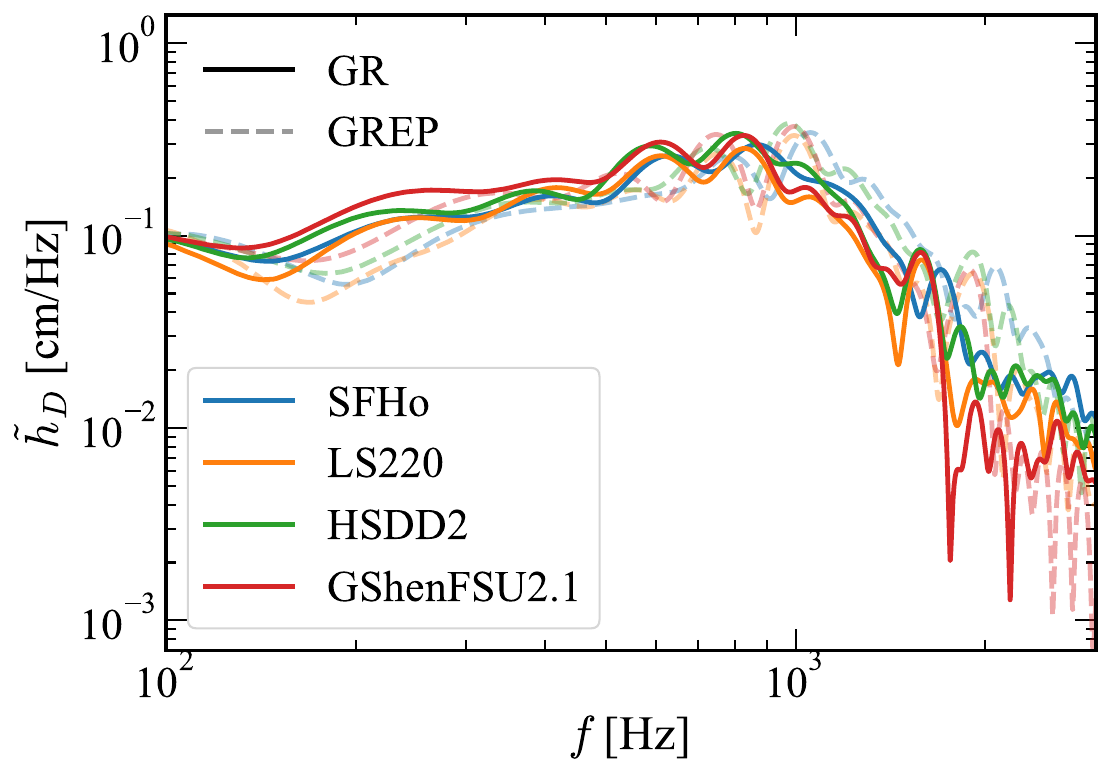} }}
    \caption{The left panel shows the gravitational wave strain as a function of time, while the right panel displays its Fourier transform for models with $T/|W| \approx 0.09$. The blue, yellow, green, and red curves represent {\tt SHFo}, {\tt LS220}, {\tt HSDD2}, and {\tt GShenFSU2.1} EOSs, respectively. Solid lines correspond to GR, and transparent dashed lines to GREP. The zero time corresponds to the time of bounce.}
    \label{fig:gw_fft_plot}
\end{figure*}

The rest of the parameters are the same in both sets of simulations. We impose axisymmetry as the stellar core remains largely axisymmetric up to $\sim \! 10 \, \mathrm{ms}$ after bounce \cite{ott:07cqg}. Magnetic fields are neglected since they do not influence the dynamics within this timescale \cite[e.g.,][]{moesta:14b}. We use 250 logarithmical radial cells spanning $3,000$ km and 40 uniform angular cells covering the upper half of the meridional plane, assuming equatorial reflection symmetry. To model neutrino processes during collapse, we use the $Y_e(\rho)$ deleptonization scheme \cite{liebendoerfer:05}. The $Y_e(\rho)$ profiles are obtained from 1D radiation-hydrodynamics calculations using {\tt GR1D} \cite{oconnor:15a}. We perform simulations for four different equations of state (EOS): {\tt SFHo} \cite{steiner:13b}, {\tt LS220} \cite{lseos:91}, {\tt GShenFSU2.1} \citep{gshen:11b}, and {\tt HSDD2} \citep{hempel:10, hempel:12}. Among the 18 EOSs in the database of \cite{richers:17}, these four EOSs were found to satisfy both observational and experimental constraints (see Fig. 1 of \cite{richers:17}). Additionally, these four EOSs produce sufficiently distinct GW signals from one another (see Fig. 10 of \cite{richers:17}).

For every EOS, we consider about 100 rotational configurations of the s12 model \cite{woosley:07}, ranging from slow to rapid rotation \cite{abdikamalov:14}. The s12 model is a red supergiant with a mass of $\sim 10.9 M_\odot$ and a radius of $\sim 4.4 \times 10^{13}$ cm, which is $\sim 630 R_\odot$. Following \cite{abdikamalov:14}, we use cylindrical rotation law
\begin{equation}
    \Omega(\varpi) = \Omega_\mathrm{0} \left[ 1+ \left( \frac{\varpi}{A} \right)^2 \right]^{-1},
\end{equation}
where $\varpi$ is the cylindrical radius, $\Omega_\mathrm{0}$ is the central angular velocity, and $A$ is a measure of differential rotation. We consider 5 different values of $A$, ranging from $300$ km, which corresponds to extreme differential rotation, to $10^4$ km, which corresponds to uniform rotation in the stellar core (see \cite{abdikamalov:14} for further details). We quantify the rotation at bounce using the parameter $T/|W|$, where $T$ is the rotational kinetic energy and $W$ is the potential binding energy. We focus on a rotation range of $0.02 < T/|W| < 0.18$. For $T/|W| < 0.02$, the rotation is too slow to cause significant quadrupole deformations, resulting in a weak bounce signal. Conversely, for $T/|W| > 0.18$, the centrifugal force becomes dominant, inhibiting the core from reaching the high densities where differences in EOS are most pronounced. In total, we have 452 GR waveforms, with 116, 120, 108, and 108 for {\tt SFHo}, {\tt LS220}, {\tt HSDD2}, and {\tt GShenFSU2.1}, respectively. For GREP, we have 412 waveforms, with 105, 105, 103, and 99 for {\tt SFHo}, {\tt LS220}, {\tt HSDD2}, and {\tt GShenFSU2.1}, respectively. The variation in the number of waveforms for each EOS arises because, as the central angular velocity increases, some EOS models do not undergo core collapse. This is due to the excessive centrifugal support present at the onset of collapse, as shown in Table III of \cite{richers:17}. As an example, Fig.~\ref{fig:gw_fft_plot} shows GWs for these EOSs in GR and GREP for models with $T/|W| \simeq 0.09$. 

The waveforms are sampled at a rate of 10 kHz, as the signal around or above this frequency does not have any significant physical component \cite[e.g.,][]{richers:17}. Moreover, at high frequencies, quantum noise becomes the dominant factor for terrestrial detectors, making it challenging to observe signals above the noise threshold \cite{martynov2016sensitivity}. We concentrate on the time interval from $-2$ ms to $6$ ms, with zero time corresponding to the bounce. This range is selected because the GW signal before $-2$ ms contains little energy, and the signal after $6$ ms includes contributions from prompt convection, which is not accurately captured in our model \cite{mitra24}. 

\begin{figure}[t]
\centering
\includegraphics[width=0.99 \linewidth]{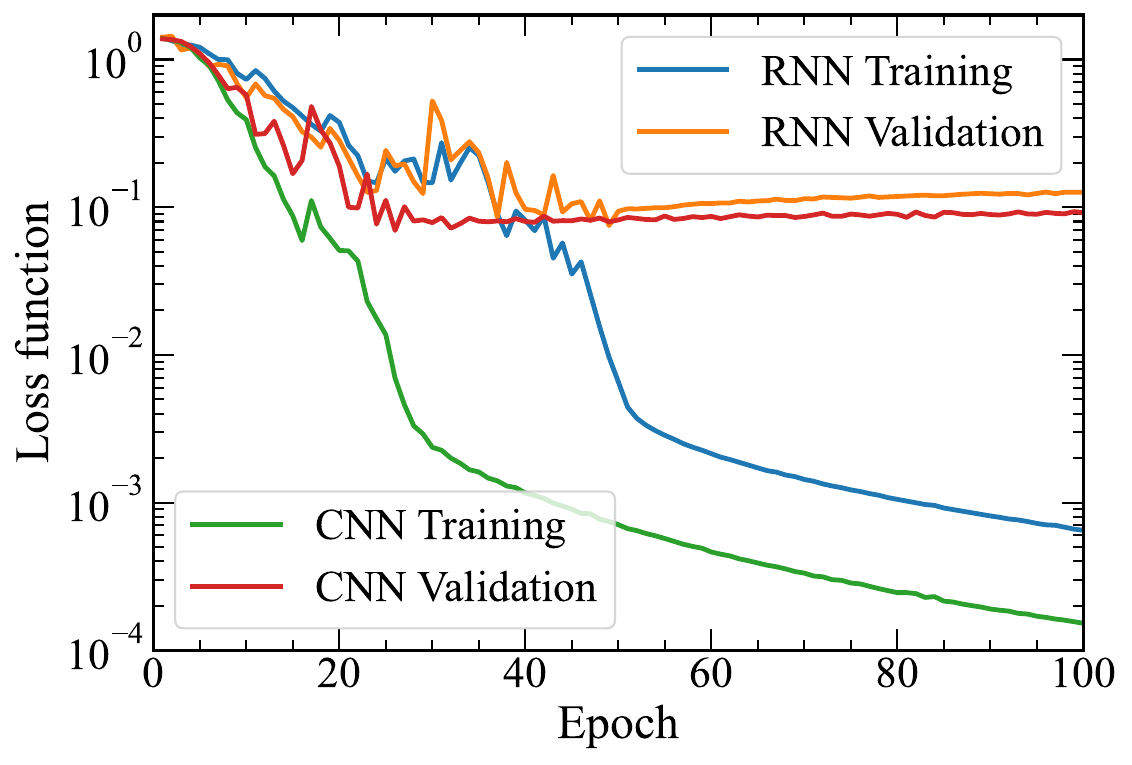}
\caption{The training and validation losses as a function of epochs are shown as blue and orange curves for the RNN, and green and red curves for the CNN. In this particular train-test split case, the minimal validation loss occurs at epoch 26 for the CNN and at epoch 49 for the RNN. After these points, the validation loss stops improving while the training loss continues to decrease, indicating overfitting.}
\label{fig:CNN_RNN_loss}
\end{figure}

\begin{table*}[t]
\caption{Architecture of the CNN model used in the study. The model consists of ten layers, starting with an Input layer, followed by three Convolutional layers, each paired with Max Pooling layers, and ending with three Dense layers after a Flatten layer. The table provides information on the type of each layer, the number of filters or units, the kernel size, the pooling size, the output shape at each layer, and the corresponding activation functions. The final Dense layer with 4 units and a Softmax activation function is used for classification.}
\vskip 0.15in
\begin{center}
\begin{small}
\begin{sc}
\begin{tabular}{C{2cm}C{3cm}C{5cm}C{2cm}C{2cm}}
\toprule
Layer & Type & Parameters & Output Shape & Activation\\
\hline
0  &  Input  &  & (81) & \\
1  &  Convolution 1D  & 32 filters, kernel size 3 & (79, 32) & ReLU \\
2  &  Max Pooling 1D  & Pool size 2               & (39, 32) &      \\
3  &  Convolution 1D  & 64 filters, kernel size 3 & (37, 64) & ReLU \\
4  &  Max Pooling 1D  & Pool size 2               & (18, 64) &      \\
5  &  Convolution 1D  &128 filters, kernel size 3 & (16, 128)& ReLU \\
6  &  Max Pooling 1D  & Pool size 2               & (8, 128) &      \\
7  &  Flatten         &                           & (1024)   &      \\
7  &  Dense           & 512 units                 & (512)    & ReLU \\
8  &  Dense           & 256 units                 & (256)    & ReLU \\
9  &  Dense           & 4 units                   & (4)      & Softmax \\
\hline
\hline
\end{tabular}
\end{sc}
\end{small}
\end{center}
\vskip 0.3in
\label{Table:CNN_Model}
\end{table*}

\begin{table*}[t]
\caption{Architecture of the RNN model employed in this study. The model consists of four layers: two SimpleRNN layers followed by two Dense layers. The SimpleRNN layers contain 64 and 128 units, respectively, while the Dense layers include 64 units with ReLU activation and 4 units with Softmax activation, facilitating the final classification output.}
\vskip 0.15in
\begin{center}
\begin{small}
\begin{sc}
\begin{tabular}{C{2cm}C{2.5cm}C{2.5cm}C{2cm}C{2cm}}
\toprule
Layer & Type      & Parameters & Output Shape & Activation\\
\hline
0  &  SimpleRNN   & 64 units  & (81, 64) &     \\
1  &  SimpleRNN   & 128 units & (128)    &    \\
2  &  Dense       & 64 units  & (64)     & ReLU     \\
3  &  Dense       & 4 units   & (4)      & Softmax  \\
\hline
\hline
\end{tabular}
\end{sc}
\end{small}
\end{center}
\vskip 0.3in
\label{Table:SimpleRNN_Model}
\end{table*}

\begin{table*}[t]
\caption{Grid of hyperparameters used for tuning classical machine learning models. The optimal hyperparameters are highlighted in bold. }
\vskip 0.15in
\begin{center}
\begin{small}
\begin{sc}
\begin{tabular}{C{6cm} | C{5.3cm} | C{5.7cm}}
\toprule
Random Forest & Support Vector Machines  & Na\"{i}ve Bayes \\
\hline
\begin{tabular}[c]{@{}l@{}} 
'n\_estimators': [50, 75, \textbf{100}, 125, 150]\\ 
'max\_depth': [\textbf{None}, 10, 15, 20]\\ 
'min\_samples\_split': [2, \textbf{5}, 10]\\ 
'min\_samples\_leaf': [\textbf{1}, 2, 4] 
\end{tabular}
& 
\begin{tabular}[c]{@{}l@{}} 
'C': [0.1, 1, \textbf{10}]\\ 
'kernel': ['linear', 'rbf', \textbf{'poly'}]\\ 
'gamma': [\textbf{'scale'}, 'auto']\\
'degree': [2, 3, \textbf{4}]
\end{tabular} 
& 
\begin{tabular}[c]{@{}l@{}} 
'var\_smoothing': [\bm{$10^{-9}$}, $10^{-8}$, $10^{-7}$,\\ \hspace{2.8cm} $10^{-6}$, $10^{-5}$] 
\end{tabular} 
\\
\hline
\hline
Logistic Regression & $k$-Nearest Neighbors  & eXtreme Gradient Boosting  \\
\hline
\begin{tabular}[c]{@{}l@{}} 
'C': [0.01, 0.1, 1, 10, \textbf{100}]\\
'penalty': [\textbf{'l1'}, 'l2', 'none']\\
'solver': ['lbfgs', 'liblinear', \textbf{'saga'}]\\
'max\_iter': [100, 200, \textbf{300}]
\end{tabular}
& 
\begin{tabular}[c]{@{}l@{}} 
'n\_neighbors': [\textbf{3}, 5, 7, 9, 11]\\
'weights': ['uniform', \textbf{'distance'}]\\
'metric': [\textbf{'euclidean'},\\ \hspace{1.5cm} 'manhattan', \\ \hspace{1.5cm} 'minkowski']\\
'p': [\textbf{1}, 2] 
\end{tabular} 
& 
\begin{tabular}[c]{@{}l@{}} 
'n\_estimators': [50, \textbf{100}, 200]\\
'max\_depth': [3, \textbf{5}, 7]\\
'learning\_rate': [0.01, 0.1, \textbf{0.2}]\\
'subsample': [\textbf{0.8}, 0.9, 1.0]\\
'colsample\_bytree': [\textbf{0.8}, 0.9, 1.0]\\
'gamma': [\textbf{0}, 0.1, 0.2]\\
'reg\_alpha': [0, \textbf{0.01}, 0.1]\\
'reg\_lambda': [1, 0.1, \textbf{0.01}]
\end{tabular} 
\\
\hline
\hline
\end{tabular}
\end{sc}
\end{small}
\end{center}
\vskip 0.3in
\label{Table:ML_grid}
\end{table*}

\subsection{Algorithms}
\label{sec:algorithm}

We use two deep learning algorithms, Convolutional Neural Networks (CNN) and Recurrent Neural Networks (RNN), and six classical ML algorithms: Random Forest (RF), Support Vector Machines (SVM), Na\"{i}ve Bayes (NB), Logistic Regression (LR), $k$-Nearest Neighbors ($k$-NN), and eXtreme Gradient Boosting (XGB). Before the ML analysis, all waveforms are normalized by dividing them by their amplitudes $D \cdot \Delta h$. We train our models and optimize their hyperparameters using the GR data.

\textbf{CNN and RNN:} We use the architecture shown in Table \ref{Table:CNN_Model} for the CNN and Table \ref{Table:SimpleRNN_Model} for the RNN. The data set is split in a ratio of 80:20, where 20\% is the test set. The remaining 80\% is then split again using an 80:20 ratio, resulting in 64\% of the original data being used as a training set and 16\% as a validation set, while maintaining the class distribution across all sets. For training, we employ the sparse categorical cross-entropy loss function, commonly used in multiclass classification, along with the Adam optimizer \cite{kingma2017adammethodstochasticoptimization}. Fig.~\ref{fig:CNN_RNN_loss} illustrates the loss function across epochs for both the CNN and RNN models.

During training, we apply the early stopping strategy to prevent overtraining. If the validation loss does not improve for 20 consecutive epochs in the case of the CNN, or for 40 consecutive epochs in the case of the RNN, we stop training and retain the model that achieved the minimum validation loss. The CNN model has three times more layers than the RNN, allowing it to learn faster and reach convergence in fewer epochs, as shown in Fig.~\ref{fig:CNN_RNN_loss}. Therefore, we set the maximum epochs to 100 for the CNN and 200 for the RNN to account for their different learning speeds.

\textbf{Classical ML algorithms:} We split our dataset with an 80:20 ratio, using the larger portion for hyperparameter tuning through the Grid Search Cross-Validation (GridSearchCV) technique \cite{gridCV}. GridSearchCV is a powerful tool in ML that optimizes model performance by systematically exploring a predefined hyperparameter space. It assesses the model's performance across various parameter combinations using 5-fold cross-validation to identify the optimal settings.

In this approach, the dataset is divided into five folds. For each combination of hyperparameters, the model is trained on four of these folds and validated on the remaining fold. This process is repeated five times, with each fold serving as the validation set exactly once. The cross-validation scores, which represent EOS classification accuracy, from these iterations are averaged to evaluate the effectiveness of each parameter combination. The combination yielding the highest average performance is selected as the optimal set of hyperparameters.

To ensure that the model with the optimal set of hyperparameters is neither overfitting nor underfitting, we compare the average accuracy on the validation folds with the average accuracy on the training folds. If the two values closely match, this indicates good generalization. However, a significant drop in validation accuracy suggests potential overfitting, prompting further adjustments.

Table \ref{Table:ML_grid} displays the hyperparameter space for the ML algorithms, with the optimal hyperparameters obtained through GridSearchCV highlighted in bold. No signs of overfitting or underfitting were observed with these optimal hyperparameters. We use this optimal set of hyperparameters in our analysis.

After determining the optimal set of hyperparameters, we split the dataset into training, validation, and test sets with a 64:16:20 ratio. Note that, for classical ML algorithms, the validation set is not utilized. We use 64$\,\%$ of the data for training and 20$\,\%$ for testing. This approach aligns with the data allocation used for CNN and RNN models, allowing for a fair comparison of performance across all algorithms.

To evaluate the performance of the EOS classification, we use accuracy, recall, and precision metrics. Where:
\begin{equation}
\text{Accuracy} = \frac{\text{Number of Correct Predictions}}{\text{Total Number of Predictions}} 
\end{equation}
and for a given EOS $i$
\begin{equation}
\text{Precision}_i = \frac{\text{TP}_i}{\text{TP}_i+\text{FP}_i} 
\end{equation}
\begin{equation}
\text{Recall}_i = \frac{\text{TP}_i}{\text{TP}_i+\text{FN}_i} 
\end{equation}
where TP$_i$, FP$_i$, and FN$_i$ represent the number of true positives, false positives, and false negatives for a given EOS $i$. Our evaluation process involves repeating the calculations 100 times, with each iteration involving a random train-test split. Repeating 100 times provides a statistically robust sample, reducing bias from any particular split. We compute the average performance across all iterations and use the standard deviation as a measure of error. This methodology helps us obtain results that are independent of specific train-test split realizations, thereby demonstrating the robustness of our findings.

\begin{table*}[t]
\caption{Classification accuracy (mean $\pm$ standard deviation) in percentage (\%) of various ML models on GR and GREP datasets. Each row shows the accuracy for the indicated dataset. GREP$\rightarrow$GR indicates training on the GREP dataset and testing on GR. GREP$^*$$\rightarrow$GR$^*$ refers to training and testing on time-normalized datasets. SVM shows the highest accuracy when trained and tested on the same dataset. All ML models trained on GREP data and tested on GR data exhibit significantly lower accuracy. Values in bold highlight the best performance for each row.}
\vskip 0.15in
\begin{center}
\begin{small}
\begin{sc}
\begin{tabular}{p{2.2cm}C{1.8cm}C{1.8cm}C{1.8cm}C{1.8cm}C{1.8cm}C{1.8cm}C{1.8cm}C{1.8cm}}
\toprule
Dataset & CNN & RNN & RF & SVM & NB & LR & $k$-NN & XGB\\
\hline
GR  & 97.4 $\pm$ 2.0  &  97.7 $\pm$ 1.9 & 96.8 $\pm$ 2.4  &  \textbf{99.5 $\pm$ 1.0} & 48.9 $\pm$ 5.0  &  95.8 $\pm$ 2.1 & 93.8 $\pm$ 2.6  &  96.6 $\pm$ 2.3 \\
GREP  & 97.2 $\pm$ 2.0  &  97.7 $\pm$ 1.8 & 96.2 $\pm$ 1.9  &  \textbf{99.1 $\pm$ 1.0} & 56.2 $\pm$ 6.6  &  97.4 $\pm$ 2.0 & 91.2 $\pm$ 3.7  &  95.9 $\pm$ 2.3 \\
GREP$\rightarrow$GR  & 37.9 $\pm$ 6.5  &  30.5 $\pm$ 5.5 & 35.2 $\pm$ 3.1  &  29.9 $\pm$ 2.5 & 38.8 $\pm$ 4.2  &  \textbf{41.4 $\pm$ 3.1} & 33.1 $\pm$ 3.5  &  34.5 $\pm$ 3.1 \\
GREP$^*$$\rightarrow$GR$^*$  & 62.0 $\pm$ 4.8  &  67.5 $\pm$ 5.2 & 43.6 $\pm$ 4.5  &  \textbf{68.0 $\pm$ 4.3} & 36.4 $\pm$ 4.7  &  57.1 $\pm$ 4.8 & 57.8 $\pm$ 4.4  &  43.6 $\pm$ 4.6 \\
\hline
\hline
\end{tabular}
\end{sc}
\end{small}
\end{center}
\vskip 0.3in
\label{Table:EOS_GR_ML}
\end{table*}

\begin{figure}
\centering
\includegraphics[width=0.95 \linewidth]{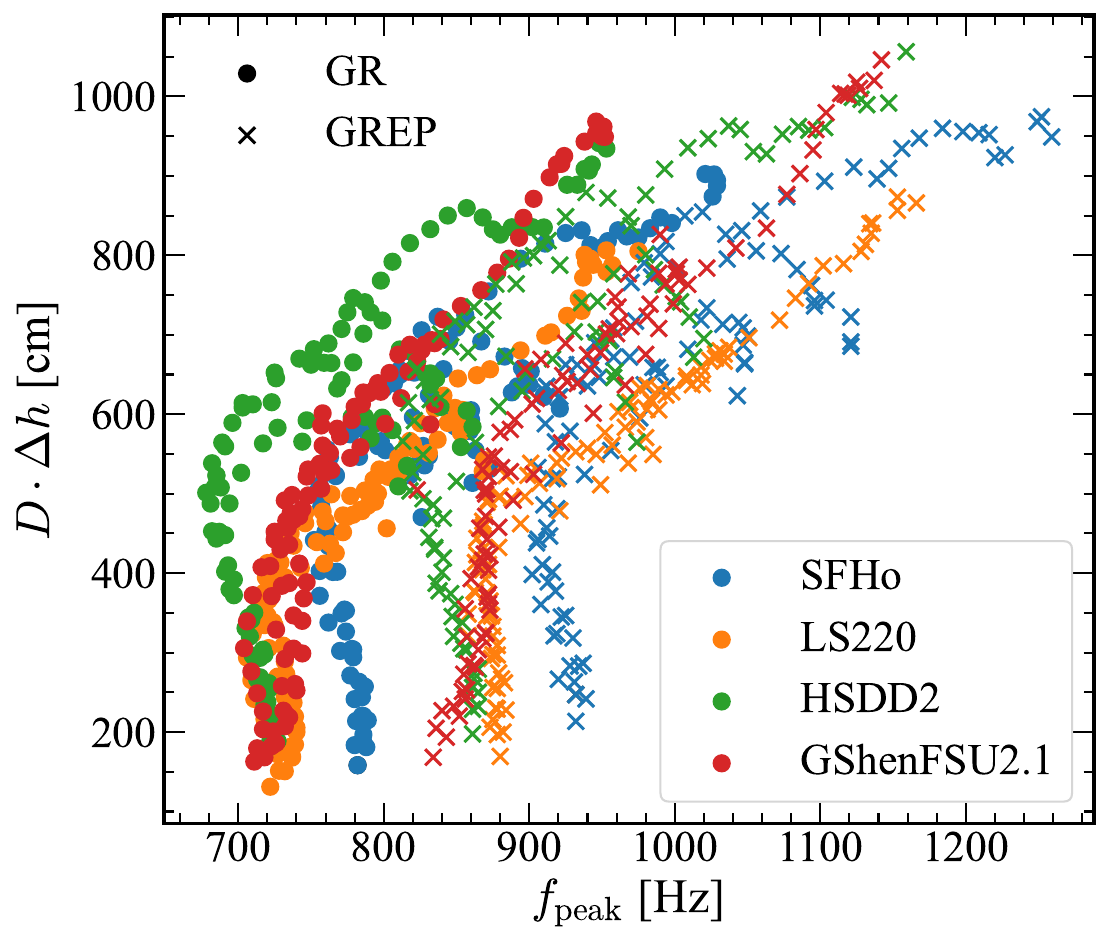}
\caption{Relationship between the amplitude $D \cdot \Delta h$ and the peak frequency $f_\mathrm{peak}$ of the bounce GW signal for different EOSs (see \cite{richers:17} for definitions of these two quantities). The circles (crosses) correspond to GR (GREP) data. For a given $D \cdot \Delta h$, the values of $f_\mathrm{peak}$ are generally higher in the GREP dataset compared to GR. This is expected since GREP lacks the time dilation effect.}
\label{fig:Dh_fpeak}
\end{figure}

\section{Results}
\label{sec:results}

Figure~\ref{fig:gw_fft_plot} shows the GW strain as a function of time for four different EOSs from $-2$ until $6$ ms after bounce. The variations between EOSs are approximately 5-10$\,\%$. Similarly, the differences between GREP and GR waveforms exhibit a 5-10$\,\%$ variation, but with a noticeably higher frequency due to the absence of time dilation in this approximation \cite{marek:06, bmueller:08}. The objective of the machine learning (ML) model is to distinguish the EOS of these signals. Unless stated otherwise, in the following we use the RF method for the time range of -2 to 6 ms, with a 10 kHz sampling rate. The study of how classification accuracy depends on the signal time range, the choice of window function, and the sampling rate is provided in Appendix ~\ref{sec:preprocessing}.

\begin{figure*}[t]
    \centering
    \subfloat{{\includegraphics[width=.45\textwidth]{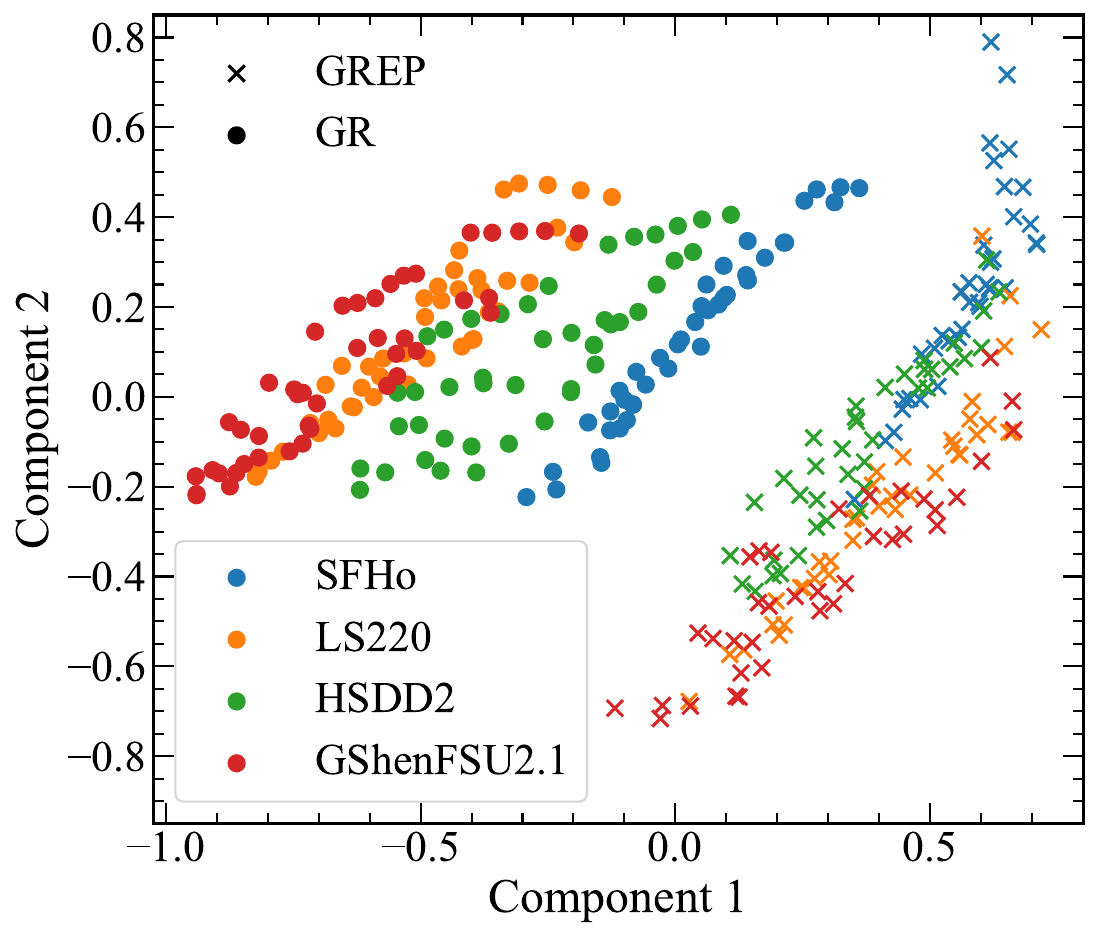} }}
    \hspace{0.05\textwidth} 
    \subfloat{{\includegraphics[width=.45\textwidth]{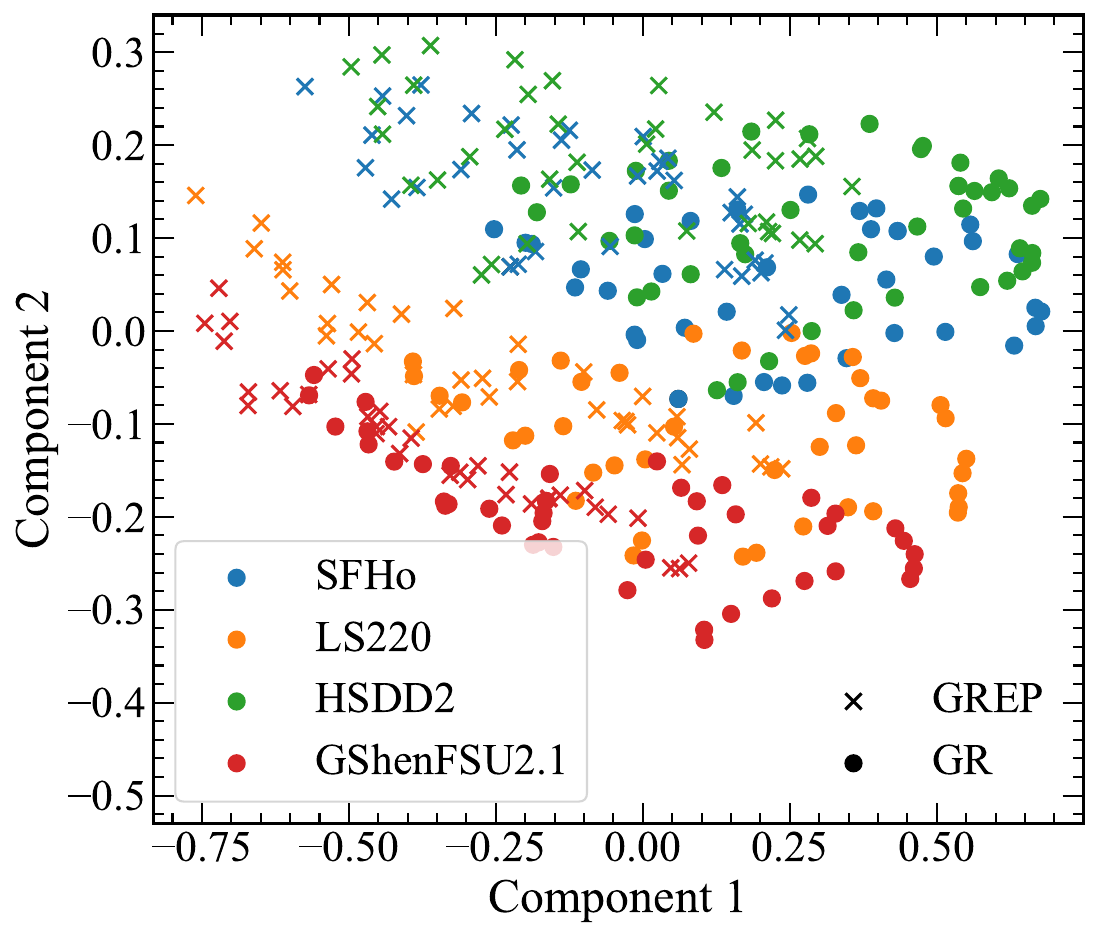} }}
    \vspace{0.02\textwidth} 
    \subfloat{{\includegraphics[width=.45\textwidth]{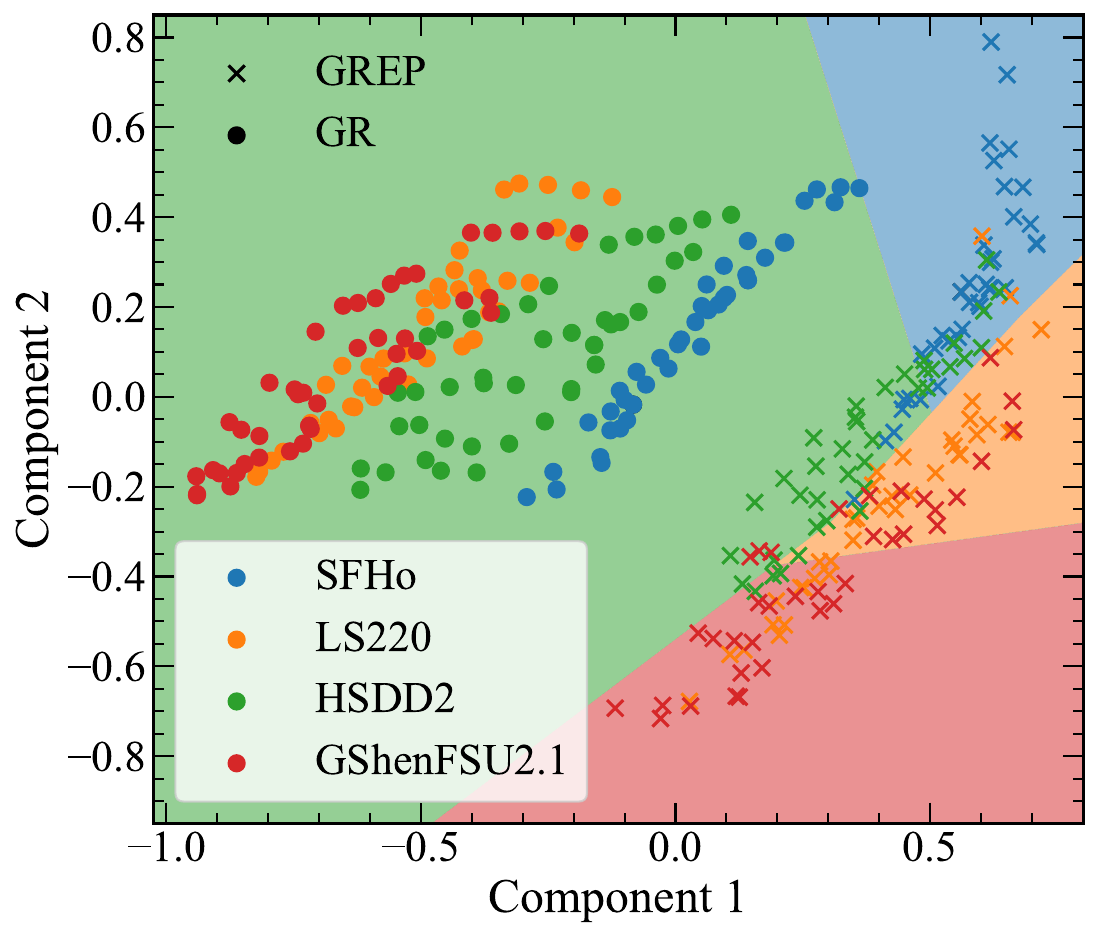} }}
    \hspace{0.05\textwidth} 
    \subfloat{{\includegraphics[width=.45\textwidth]{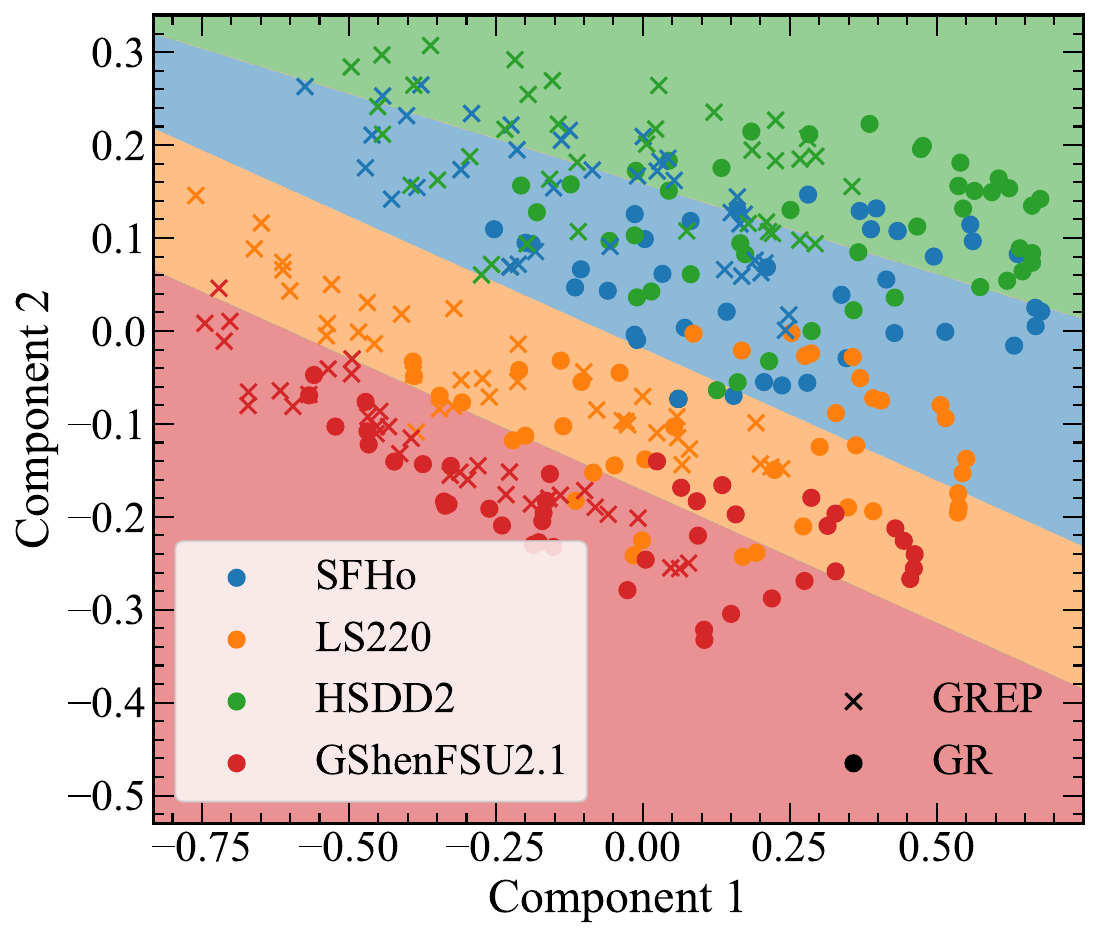} }}
    \caption{{Visualization of the first two principal components of the GR and GREP datasets. The top panels display the distribution of waveforms for the original dataset (left) and the time-normalized dataset (right). In the left panel, the GR data (circles) and GREP data (crosses) are clearly separated and form distinct groups. However, in the right panel, after time normalization, this distinction fades, and the data points cluster based on the equation of state (EOS), rather than the method (GR or GREP). The bottom panels show the decision boundaries formed by the SVM algorithm, with GREP used for training and GR for testing. The colored regions (blue for {\tt SFHo}, orange for {\tt LS220}, green for {\tt HSDD2}, and red for {\tt GShenFSU2.1}) indicate the classification regions.}}
    \label{fig:PCA}
\end{figure*}

\subsection{ML models}
\label{sec:ML_models}

Table \ref{Table:EOS_GR_ML} presents the classification accuracy results across different ML models. Each entry in the table represents the mean accuracy and the corresponding standard deviation.

The SVM model consistently outperforms other models, achieving the highest mean accuracy of $99.5 \pm 1.0\,\%$. The strong performance of the SVM model is likely due to its configuration and the nature of the GW data. The polynomial kernel with a degree of 4 allows the SVM to capture complex, non-linear relationships necessary for classifying different EOSs. The regularization parameter $C=10$ effectively balances the decision boundary margin with a controlled level of misclassification, enhancing the model’s ability to generalize without overfitting. These factors together likely contribute to the SVM model's high classification accuracy. 

Along with SVM, both RNN and CNN achieve strong performance, with average accuracies surpassing 97$\, \%$. RF, XGB, LR, and k-NN exhibit slightly lower but still high average accuracies of 96.8$\,\%$, 96.6$\,\%$, 95.8$\,\%$, 93.8$\,\%$, respectively (cf. Table~\ref{Table:EOS_GR_ML}). In contrast, Na\"{i}ve Bayes shows a significantly lower accuracy of $48.9 \pm 5 \, \%$, primarily due to its assumption of feature independence, which overlooks the correlations and temporal patterns that exist in time series data. Overall, except for Na\"{i}ve Bayes, all models demonstrate strong performance in this classification task.

\begin{figure*}
    \centering
    \subfloat{{\includegraphics[width=.45\textwidth]{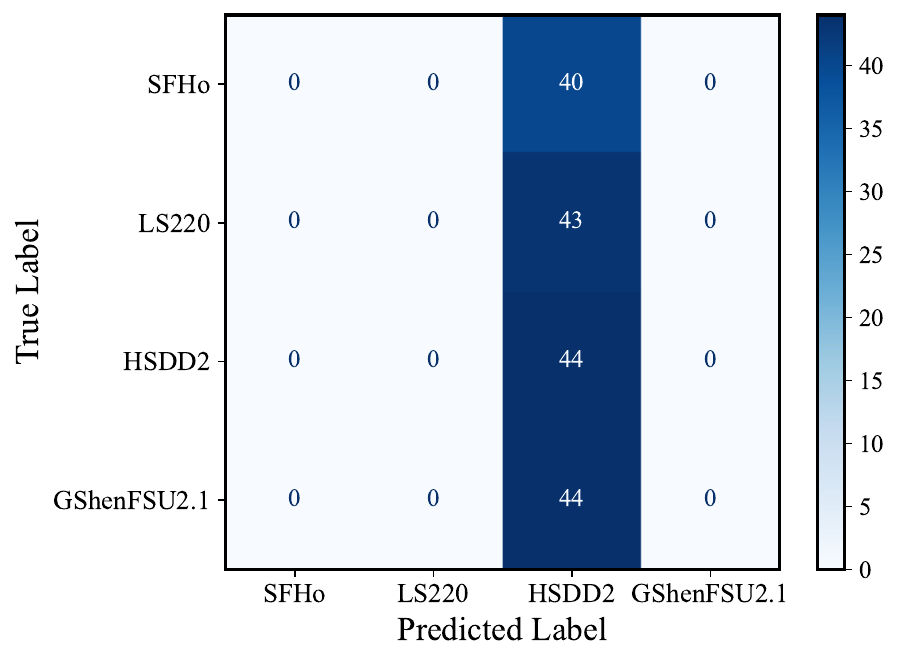} }}
    \hspace{0.05\textwidth}
    \subfloat{{\includegraphics[width=.45\textwidth]{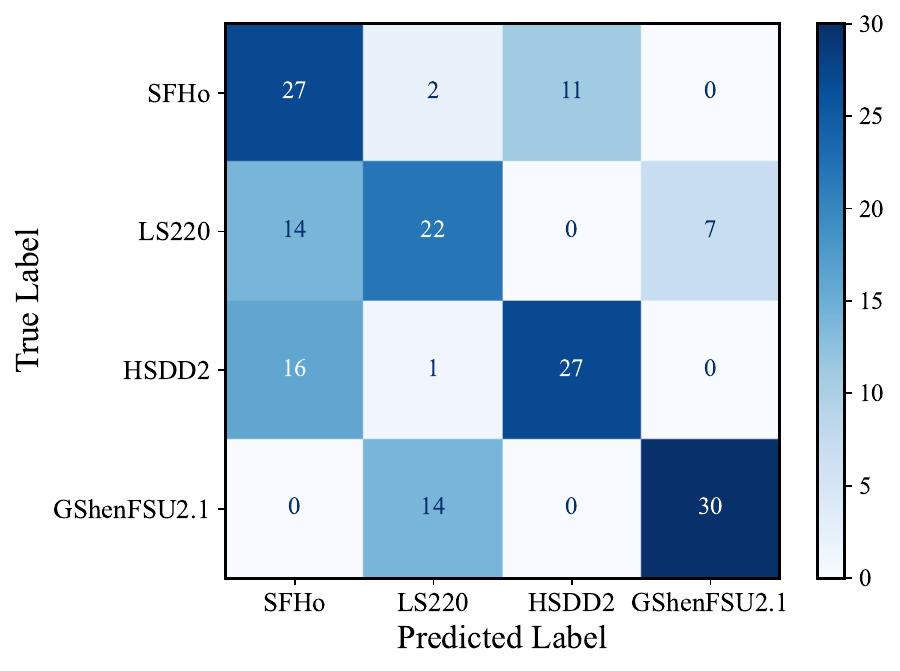} }}
    \caption{{Confusion matrices illustrating the performance of the SVM algorithm in classifying GR data when trained on GREP data. The left panel shows the confusion matrix for the original dataset. The right panel displays the confusion matrix after time normalization. In the original dataset, substantial misclassifications, particularly as {\tt HSDD2}, result in lower classification accuracy of $\sim 25\,\%$. After time normalization, accuracy improves to $\sim 60\,\%$.}}
    \label{fig:Confusion_matrix}
\end{figure*}

\subsection{GR vs effective potential}
\label{sec:gr_vs_grep}

Next, we assess the ability of ML models trained on GREP data to classify the EOS from realistic GW signals, which are obtained from GW simulations. While all ML methods discussed in this paper were applied, for brevity and the reasons outlined below, we focus on the results obtained with the SVM method.

The model trained and tested on GREP data achieves an accuracy of $99.1 \pm 1.0 \, \%$, which is comparable to the $99.5 \pm 1.0 \, \%$ accuracy obtained for the model trained and tested on GR data. However, when the model trained on GREP data is used to classify the GR data, the accuracy drops to $29.9 \pm 2.5 \, \%$. This is expected, as GREP and GR waveforms differ by $\sim$ 5-10$\,\%$, which is of the same order as the difference between different EOSs, as can be gleaned from  Fig.~\ref{fig:gw_fft_plot}. For this reason, the model trained on GREP struggles to classify GR signals. 

Another factor contributing to the lower accuracy is the frequency difference. Since GREP lacks the time dilation effect, it produces dynamics with higher frequencies \cite{bmueller:08}. This is evident in Fig.~\ref{fig:Dh_fpeak}, which plots the GW amplitude $D\cdot \Delta h$ against the peak frequency $f_\mathrm{peak}$ for both GR and GREP datasets (see \cite{richers:17} for definitions of these two quantities). As we can see, the GREP waveforms have $\sim 15 \, \%$ higher frequencies compared to the GR models. 

To gain deeper insight, we design a simple experiment. We select waveforms with $T/|W|$ below 0.06, where the amplitude-normalized waveforms are similar to each other \cite{richers:17}. We then generate another dataset by normalizing time by $f_\mathrm{peak}$, as done in \cite{pastor24}. Next, we apply Principal Component Analysis (PCA) to reduce the dimensionality of the waveforms in both datasets from 81\footnote{For our time series data, a 10 kHz sampling rate corresponds to a 0.1 ms step size, resulting in 81 data points over the range [\textendash2, 6] ms.} to 2. We then analyze the resulting data.

The upper panel of Fig.~\ref{fig:PCA} shows each waveform as the first two principal components. The left panel corresponds to the original dataset, while the right panel corresponds to the time-normalized dataset. In the left panel, the GR and GREP data points are distinctly grouped and clearly separated from each other. In contrast, the right panel demonstrates that after the temporal normalizing, the distinction between GR and GREP becomes less pronounced and the data points cluster into somewhat distinct groups based on their EOSs.

We apply the SVM algorithm with a linear kernel to gain a deeper understanding of the impact of the temporal normalization. The blue, orange, green, and red regions in the bottom panels of Fig.~\ref{fig:PCA} correspond to the {\tt SFHo}, {\tt LS220}, {\tt HSDD2}, and {\tt GShenFSU2.1} EOSs, respectively. These regions represent the decision boundaries formed by the SVM model. Any point within this region will be classified according to the EOS associated with that region. Most GREP points fall within the decision boundaries of their respective EOSs. In contrast, all GR points are located within the regions corresponding to the {\tt HSDD2} EOS for the GREP data, meaning all GR signals will be classified as {\tt HSDD2}. However, the time-normalized GR data points, shown in the right panel, align better with the decision boundaries. Thus improves classification accuracy to $\sim 60 \, \%$. Despite this, the accuracy remains substantially lower than the $99.5 \pm 1.0 \, \%$ achieved when using GR data for both training and testing.

The left and right panels of Fig.~\ref{fig:Confusion_matrix} show the corresponding confusion matrices for the original and time-normalized data. For the original data, all GR points fall within the {\tt HSDD2} EOS region of the GREP data, leading to their classification as {\tt HSDD2} (left panel). The precision and recall for {\tt HSDD2} are $\sim$ 0.25 and 1, respectively. For all other EOSs, precision is undefined due to the division by zero, while recall is equal to zero. For the time-normalized data (right panel), {\tt GShenFSU2.1} has the highest recall and precision, making it the best-classified EOS, while {\tt LS220} performs the worst. Notably, {\tt SFHo} is often misclassified as {\tt HSDD2} and {\tt LS220}. 

To verify the robustness, we conduct a similar test on the full data without applying the PCA reduction. We achieve a comparable accuracy. The accuracy of all ML models is provided in Table \ref{Table:EOS_GR_ML}. These results indicate that the time normalization partially mitigates the absence of time dilation effects in GREP, leading to improved classification accuracy. However, even the best-performing model, SVM, reaches only $68.0 \pm 4.3 \, \%$ accuracy. 

Strictly speaking, since $f_\mathrm{peak}$ encapsulates some EOS-dependent information \cite{richers:17}, time normalization by $f_\mathrm{peak}$ is expected to weaken the distinctions between GWs corresponding to different EOSs. To assess the impact of this effect, we conduct tests on models trained and tested using either GR or GREP data (cf. Section \ref{sec:time_norm} for details). Our analysis shows that the resulting drop in accuracy due to this effect is $\sim 6 \%$, which is insufficient to account for the low EOS classification accuracy observed in GREP-trained models. This suggests that GREP waveforms lack the precision required to capture subtle EOS-related features. 

\section{Conclusion}
\label{sec:concl}

In this study, we evaluated the impact of various machine learning models, parameter configurations, and data preprocessing techniques on the accuracy of equation-of-state (EOS) classification using bounce gravitational wave (GW) data from rotating core-collapse supernovae. 

As expected, we find that the signal length has a significant impact on accuracy: longer signals provide more information, enhancing the model's ability to classify accurately (cf. Section~\ref{sec:time} for details). Applying a Tukey window has little impact on classification accuracy, as most EOS-sensitive information is concentrated in the middle part of the signal, while the window primarily affects the signal's tails (cf. Section~\ref{sec:window}). We find that the sampling frequency does not influence accuracy as long as it exceeds 6 kHz or above, which is unsurprising since the GW signal lacks substantial components at these or higher frequencies (cf. Section~\ref{sec:sampling} for details).

Our results show that Support Vector Machines (SVM), Recurrent Neural Networks (RNN), and Convolutional Neural Networks (CNN) achieve the highest classification accuracy, exceeding 97$\,\%$ for a set of four EOSs. Random Forest (RF), XGBoost (XGB), Logistic Regression (LR), and $k$-Nearest Neighbors ($k$-NN) also perform well, with average accuracies of 96.8$\,\%$, 96.6$\,\%$, 95.8$\,\%$, and 93.8$\,\%$, respectively. In contrast, the Na\"{i}ve Bayes method demonstrates the lowest accuracy, falling below 50$\,\%$ (cf. Section~\ref{sec:ML_models} for details).

We also assess the impact of approximating the gravitational wave (GW) signal using the general relativistic effect potential (GREP) on classification performance. Models trained on GREP data classify waveforms based on general relativity (GR) with an average accuracy of approximately 35$\,\%$, which is significantly lower than the over 90$\,\%$ accuracy achieved by models trained on GR signals. This outcome is expected, as GREP can approximate GR gravity but does not capture other relativistic effects, such as time dilation, leading to a generally higher frequency in GREP. However, when the GW signal is normalized by the peak GW frequency, the classification accuracy improves substantially, but still remains below 70$\,\%$ (cf. Section~\ref{sec:gr_vs_grep} for details). This suggests that the GREP approximation produces waveforms lacking the precision necessary to capture subtle signal characteristics, such as the EOS. 

Our work has a number of limitations. In particular, we consider only one progenitor model. While different progenitors are expected to produce similar bounce GW signals for the same angular momentum distribution \cite{ott12correlated, mitra23}, variations of a few percent can still occur. Moreover, we do not include detector noise. These factors may affect the ability of ML models to identify the EOS. The presence of detector noise reduces classification accuracy. However, since the primary aim of this work is to compare the relative performance of ML models, this comparison is unlikely to be significantly impacted by noise. Additionally, our conclusion that the GREP approximation is too inaccurate for EOS classification remains unaffected by the noise. Beyond EOS classification, our ultimate goal is to extract nuclear matter parameters through regression analysis using a parameterized family of EOSs. Finally, we have not incorporated traditional frequentist or Bayesian statistical methods in this study. We will address these issues in more detail in future work. 

\begin{acknowledgments}

The authors would like to thank the anonymous referees for their insightful comments and suggestions, which improved the quality of this manuscript. This research is supported by the Science Committee of the Ministry of Science and Higher Education of the Republic of Kazakhstan (Grant No. AP19677351 and AP13067834) and by the Nazarbayev University Faculty Development Competitive Research Grant Program, with Grant No. 11022021FD2912 and 201223FD8821. 

\end{acknowledgments}

\section*{DATA AVAILABILITY}
The gravitational waveforms are publicly accessible at \href{https://doi.org/10.5281/zenodo.13774509}{https://doi.org/10.5281/zenodo.13774509}.

\appendix

\section{Dependence on data pre-processing}
\label{sec:preprocessing}

In this section, we examine how classification accuracy is influenced by the signal's time range, the choice of window function, the sampling rate, and the time normalization. 

\subsection{Time range}
\label{sec:time}

\begin{figure}
\centering
\includegraphics[width=0.95 \linewidth]{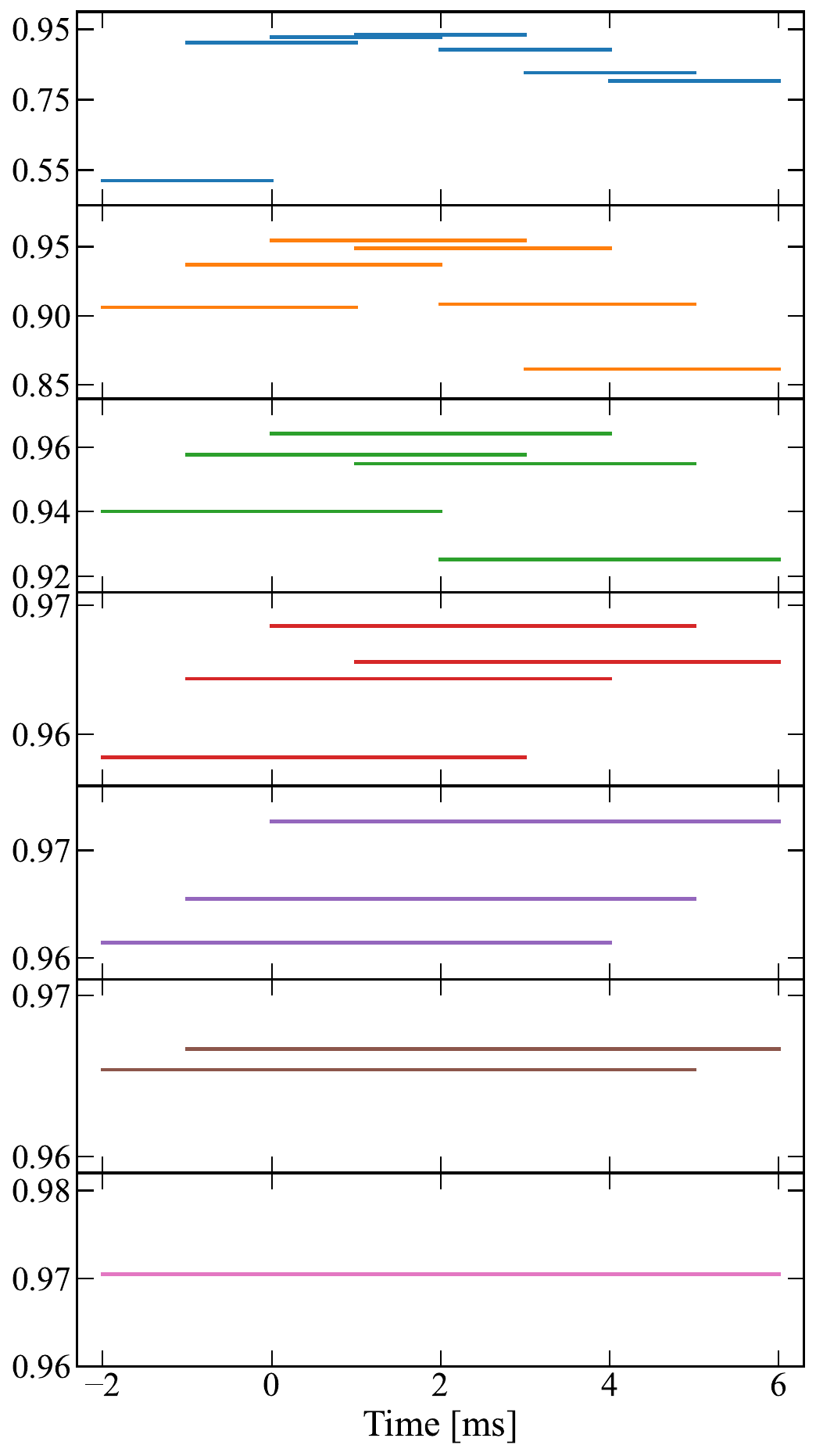}
\caption{Classification accuracy of the Random Forest (RF) model across various signal lengths and locations, using a sliding window approach. Each signal is represented as a horizontal line, with its start and end points marking the signal's duration. The sliding window progresses in 1 ms increments. The vertical position of each line corresponds to the classification accuracy value.}
\label{fig:Time_range}
\end{figure}

\begin{figure}
\centering
\includegraphics[width=0.95 \linewidth]{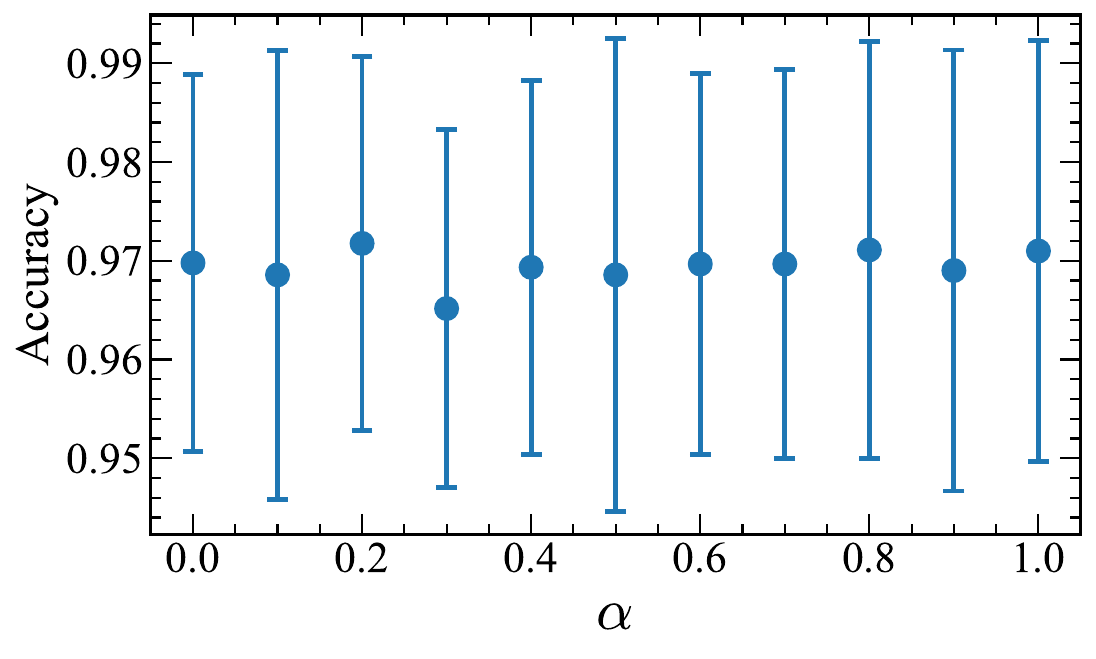}
\caption{Classification accuracy of the EOS model as a function of the tapering parameter of the Tukey window. The dots represent the mean accuracy, and the error bars indicate the standard deviation.}
\label{fig:Window}
\end{figure}

\begin{figure}
\centering
\includegraphics[width=0.95 \linewidth]{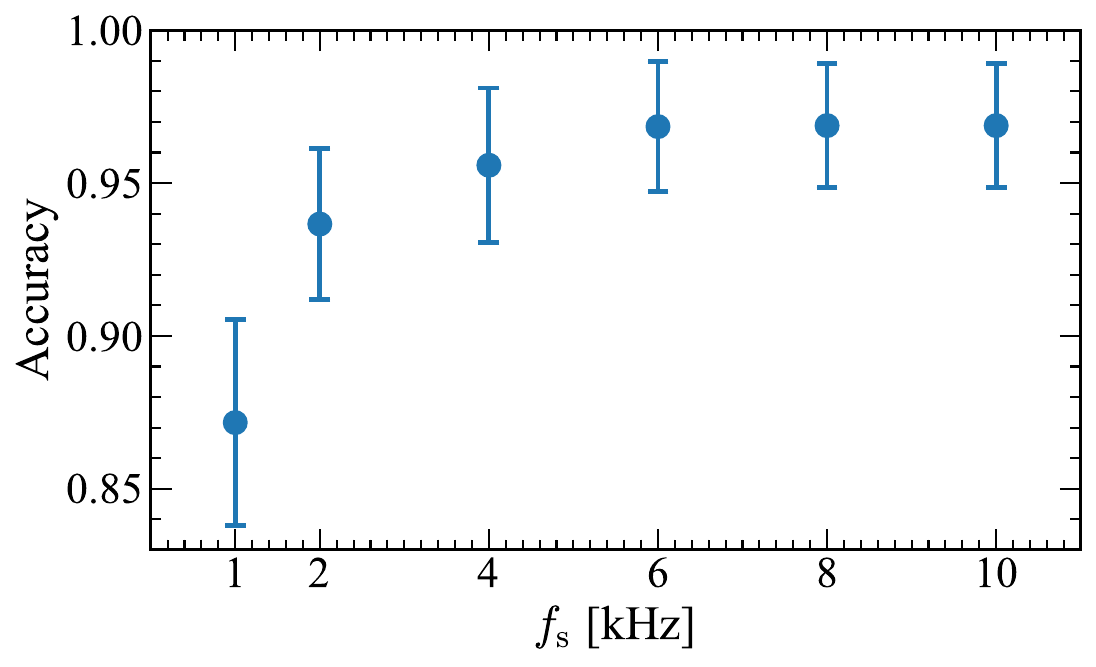}
\caption{Classification accuracy of the EOS model as a function of the signal’s sampling frequency. The dots represent the mean accuracy, and the error bars indicate the standard deviation.}
\label{fig:freq}
\end{figure}

We first study the influence of the signal length and duration on the accuracy of the classifier using a sliding window approach. We consider signal lengths from 2 to 8 ms, with each length increasing in 1 ms increments. For each signal length, a sliding window moves across the entire signal in 1 ms steps. This is shown in Fig.~\ref{fig:Time_range}, where each horizontal line represents a signal length and location in units of ms. The vertical position of each line corresponds to the classification accuracy value.

As we can see, accuracy varies with both signal length and location. For shorter signals, up to 5 ms, accuracy is more sensitive to the position of the window. This is because shorter windows do not capture the whole region that contains all key information. For example, for the signal from -2 to 0 ms, the classification accuracy is below 55$\,\%$, while for region from 0 to 2 ms, it raises to about 93 \%. The highest accuracy of 94$\,\%$ is achieved between 1 and 3 ms (cf. top panel of Fig.~\ref{fig:Time_range}). This suggests that PNS oscillations in the early post-bounce phase carry the most inferrable information about the EOS.

In contrast, longer signals are less affected by window position, resulting in relatively stable accuracy across different signal locations. For the signal with the widest range from $-2$ to $6$ ms, we obtained accuracy of $96.8 \pm 2.4 \, \%$ (cf. bottom panel of Fig.~\ref{fig:Time_range}). Note that this accuracy exceeds the 87$\,\%$ reported in our previous work \cite{mitra24}, as here we focus on waveforms with $0.02 < T/|W| < 0.18$, where distinguishing the EOS is easier.

In the following analysis, we use the entire signal from $-2$ to $6$ ms to ensure that the classifier has access to the maximum amount of information.

\begin{table*}[ht!]
\caption{Classification accuracy (mean $\pm$ standard deviation) in percentage (\%) of various ML models on GR and GREP time-normalized datasets.}
\vskip 0.15in
\begin{center}
\begin{small}
\begin{sc}
\begin{tabular}{p{2.2cm}C{1.8cm}C{1.8cm}C{1.8cm}C{1.8cm}C{1.8cm}C{1.8cm}C{1.8cm}C{1.8cm}}
\toprule
Dataset & CNN & RNN & RF & SVM & NB & LR & $k$-NN & XGB\\
\hline
GR$^*$  & \textbf{94.1 $\pm$ 3.2}  &  93.9 $\pm$ 3.5 & 90.7 $\pm$ 3.5  &  93.3 $\pm$ 2.5 & 47.0 $\pm$ 5.3  &  89.9 $\pm$ 2.8 & 80.8 $\pm$ 3.7  &  93.2 $\pm$ 3.2 \\
GREP$^*$  & \textbf{94.9 $\pm$ 3.3}  &  91.2 $\pm$ 3.9 & 88.8 $\pm$ 4.0  &  93.0 $\pm$ 2.8 & 47.0 $\pm$ 5.7  &  92.4 $\pm$ 2.7 & 78.7 $\pm$ 3.9  &  90.5 $\pm$ 3.2 \\
\hline
\hline
\end{tabular}
\end{sc}
\end{small}
\end{center}
\vskip 0.3in
\label{Table:norm table}
\end{table*}

\subsection{Window function}
\label{sec:window}

Edwards (2017) \cite{edwards2017} used a Tukey window to mitigate spectral leakage after downsampling the GW signal. In this section, we investigate the impact of the window on the classification accuracy. Fig.~\ref{fig:Window} shows the classification accuracy for varying $\alpha$, the tapering parameter of the Tukey window. As we can see, the accuracy is insensitive to $\alpha$. This can be attributed to the fact that, as indicated by the top panel in Fig.~\ref{fig:Time_range}, all key information for EOS classification is concentrated in the middle part of the signal. Since the Tukey window primarily affects the tails, it has little impact on the classification accuracy.

\subsection{Sampling rate}
\label{sec:sampling}

Figure \ref{fig:freq} shows the average classification accuracy as a function of the sampling frequency $f_\mathrm{s}$ of the signal. When the frequency exceeds $ 6 \, \mathrm{kHz}$, the classification accuracy stabilizes at around $0.97$, becoming largely insensitive to further increases. This behavior is expected, as GW signals lack significant physical content above $\sim 6 \, \mathrm{kHz}$ \cite[e.g.,][]{richers:17}. This is consistent with our previous work \cite{mitra24}, where a similar stabilization in accuracy was observed for sampling rates of 100 kHz, 10 kHz, and 5 kHz. However, for frequencies below $\sim 4 \, \mathrm{kHz}$, the classification accuracy gradually declines, reaching about $ 0.87$ at $1 \, \mathrm{kHz}$, suggesting that some of EOS-specific information lies outside this sampling rate.

\subsection{Time normalization}
\label{sec:time_norm}

In this section, we investigate the effect of time normalization on the EOS classification accuracy. We consider two cases: (1) models trained and tested on GR data and (2) models trained and tested on GREP data. The normalization is performed by scaling the time by the peak frequency $f_\mathrm{peak}$ of the GW signal. Table \ref{Table:norm table} provides the classification accuracy results across different ML models. Compared to the results without time normalization (provided in Table~\ref{Table:EOS_GR_ML}), time normalization results in a $\sim 6\,\%$ drop in the EOS classification accuracy. The reason for this drop is simple: $f_\mathrm{peak}$ encapsulates some EOS-dependent information \cite{richers:17}. As a result, normalizing time by this factor reduces the distinctions between the GW signals corresponding to different EOSs.

\section{General relativistic effective potential}
\label{sec:grep}

The GR effective potential is obtained by replacing the spherically symmetric part ${\bar \Phi}$ of the Newtonian potential $\Phi$ with the relativistic Tolman-Oppenheimer-Volkoff (TOV) solution ${\bar \Phi}_\mathrm{TOV}$ \citep{keil:97, Rampp02Radiation}:
\begin{equation}
    \label{eq:phi}
    \Phi_\mathrm{GREP} = \Phi - {\bar \Phi} + {\bar \Phi}_\mathrm{TOV}.
\end{equation}
The TOV potential is obtained as
\begin{equation}
    {\bar \Phi}_\mathrm{TOV} = \int_\infty^r 
    \left[ {m}_\mathrm{TOV} + 4 \pi r'^3 \left (P + P_\nu \right) \right]
    \frac{h}{\Gamma^2} \frac{dr'}{r'^2},
\end{equation}
where $h=1+\epsilon + P / \rho$ is the specific enthalpy, $\epsilon$ is the specific internal energy,  $P$ is pressure, and $P_\nu$ is the neutrino pressure. Parameter $\Gamma$ is given by
\begin{equation}
\label{eq:gamma}
    \Gamma = \left(1+\upsilon_r^2 - \frac{2 m_\mathrm{TOV}}{r} \right)^{0.5},
\end{equation}
where $\upsilon_r$ is the angle-averaged radial velocity. Following \cite{marek:06}, we define the enclosed mass as  
\begin{equation}
\label{eq:mtov}
    m_\mathrm{TOV} = 4 \pi \int_0^r  \left( \rho + \rho \epsilon  + E_\nu + \frac{\upsilon_i F^i_\nu}{\Gamma}  \right) \Gamma r'^2 dr'.
\end{equation}
This choice of $m_\mathrm{TOV}$, referred to as "case A", was found to better reproduce exact GR results \citep{marek:06}. Here, $E_\nu$ and $F_\nu$ and the energy and momentum density of neutrinos. Since the $Y_e(\rho)$ deleptonization scheme we employ does not explicitly evolve the neuntiro momentum density \citep{liebendoerfer:05}, we set it to zero in both GR and GREP versions of the code. Since parameter $\Gamma$ depends on $m_\mathrm{TOV}$ and vice versa, we can obtain these quantities either by solving equations (\ref{eq:gamma}) and (\ref{eq:mtov}) iteratively \citep{oconnor18two} or through ODE integration \citep{bmueller:08}. 

This method relies on the spherically symmetric component of the GR potential, but because rotation breaks this symmetry, its accuracy may decrease with increasing rotation. Rotational corrections, as suggested by \citet{bmueller:08} and recently used in \citet{mueller:20b, varma23}, can partially mitigate the issue. However, these corrections have not been implemented in our work, as our goal is to investigate the most commonly used formulation of GREP found in the literature.

\nocite{*}

\bibliography{apssamp}

\end{document}